\documentclass[final]{ws-ijmpd}

\def\Box{\nabla^2}
\def\al{\alpha}
\def\be{\beta}

\def\La{\Lambda}
\def\na{\nabla}

\def\si{\sigma}

\def\Ga{\Gamma}

\def\La{\Lambda}

\def\beq{\begin{eqnarray}}
\def\eeq{\end{eqnarray}}

\begin{document}

\markboth{Ana Pelinson }{Revisiting the modified Starobinsky model
with a cosmological constant}

%
\catchline{}{}{}{}{}
%

\title{Revisiting the modified Starobinsky model with cosmological constant}

\author{Ana Pelinson}

\address{High Energy Physics Group, Dept. ECM, and Institut de Ci{\`e}ncies del Cosmos\\
    Univ. de Barcelona, Av. Diagonal 647, E-08028 Barcelona, Catalonia, Spain\\
ana.pelinson@gmail.com}

\maketitle

\begin{history}
\received{Day Month Year}
\revised{Day Month Year}
\comby{Managing Editor}
\end{history}

\begin{abstract}
The Starobinsky model is a natural inflationary scenario in which inflation arises due to quantum effects of  the massless matter fields. A modified version of the  Starobinsky (MSt) model takes the masses of matter fields and the cosmological constant, $\Lambda$, into account. The equations of motion become much more complicated however approximate analytic and numeric solutions are possible. In the MSt model, inflation starts due to the supersymmetric (SUSY) particle content of the underlying theory and the transition to the radiation dominated epoch occurs due to the relatively heavy s-particles decoupling. For $\Lambda=0$ the inflationary solution is stable until the last stage, just before decoupling. In the present paper we generalize this result for $\Lambda\neq 0$, since $\Lambda$ should be non-vanishing at the SUSY scale.  We also take into account the radiative corrections to $\Lambda$. The main result is that the inflationary solution of the MSt model remains robust and stable.
\end{abstract}

\keywords{effective action; inflation; conformal anomaly; supersymmetry.}

\section{Introduction}

The standard cosmological model covers a wide class of
phenomena and fits the current observational tests with great
success. However, this model has problems of\,\cite{frid,grav} the initial singularity,
horizon, flatness and monopoles in the early period of the
universe. These problems can be solved if we
assume that the primordial universe starts with a very fast
expansion, denominated inflation by Guth in 1981.\cite{Guth}

An essential natural inflationary scenario is
one in which inflation is driven by quantum corrections to the
Einstein-Hilbert action, suggested by Starobinsky in 1980.\cite{star}
The Starobinsky model is based on the
semiclassical approach to quantum field theory (QFT) in curved
space-time.
Within this theory the metric is treated as a
classical background for the quantum dynamics of the matter
fields.
This approach presents a consistent theory at
energies of a few orders of magnitude below the Planck scale.\cite{birdav,book}

In the original Starobinsky model, inflation is a consequence of the
quantum effects of massless matter fields.\cite{star} The model
assumes a non-minimal conformal coupling between the scalar field
and gravity, $\xi=1/6$. In this case, the massless matter fields
are conformally invariant having a traceless stress tensor at the
classical level. However, the one-loop contributions create
a trace anomaly which changes the dynamics of the conformal factor
of the metric (see Refs. \refcite{star,fhh,vile}) and also the
metric and density
perturbations.\cite{much,star7982,star83} An alternative option is to
apply the effective action method, using the conformal anomaly to
calculate the induced effective action.\cite{book,rie}\cdash\cite{mott}
Inflation naturally arises from the total action which is obtained
from the sum of the anomaly-induced effective action to the classical terms,
including the Einstein-Hilbert one.\cite{buodsh85}\cdash\cite{
wave}

An alternative version of the  Starobinsky model was proposed in
Refs.~\refcite{graceexit,massivecase,asta}. The main advantage of
this modified
version is that inflation starts in the stable regime
which is afterwords interpolating to an
unstable regime at the end of inflation.\cite{star,vile,anju}
The modified
Starobinsky (MSt) model is a natural extension of the Starobinsky
model. In the MSt version, inflation is due to the contribution
of the quantum effects of both massless conformal and
massive matter fields.\cite{graceexit}\cdash\cite{asta}
The massive theory is not
conformally invariant at the classical level due to the masses of
the scalar and fermion fields. However, using a conformal
description, the massive matter fields
become conformally invariant and we can use the conformal anomaly
method to derive the effective action.\cite{massivecase,asta,cosmon1}

The stability condition depends on the particle content
of the underlying quantum field theory.
Assuming supersymmetry in the high energy region, the
supersymmetric (SUSY) particle content provides an initial
inflationary period stable under small perturbations of the
conformal factor.\cite{asta} The Hubble parameter, $H$, is not
constant at this stage as in the original Starobinsky model.
Instead, the inflationary expansion
is slowing down due to the contributions of massive
particles. At some point the stable inflation becomes
unstable when the s-particles
decouple and supersymmetry breaks down.

For a vanishing cosmological constant (CC) we showed
that the numerical solution of the MSt model has an accurate
approximate analytic solution which is robust during the entire inflationary stage.\cite{massivecase,asta}
Recent astronomical
observations indicate a small value for the CC at present, $\Lambda_{\rm present}
\approx 0.7 \rho_c^0 \sim 10^{-121}\,M_{\rm Pl}^4\,$. There is a
large discrepancy comparing $\Lambda_{\rm present}$ with the vacuum energy
density $\rho_{\rm vac}\equiv \Lambda/(8\pi G)$ at higher energy
scales, where $M_{\rm Pl} = 1/\sqrt{8\pi G} = 2.44 \times 10^{18}
\, {\rm GeV}$ is the reduced Planck mass. This discrepancy is
known nowadays as the old cosmological problem (see Ref.~\refcite{weinb}
for a classical review). In fact, $\rho_{\rm vac}\sim M_{\rm
SUSY}^4\simeq 10^{-12}M_{\rm Pl}^4 $
for example, on the SUSY scale.
The value of $\rho_{\rm vac}$ is due to the extremely exact
fine-tuning of the vacuum counterpart of the CC today and to
the abrupt change of the CC due to the induced counterpart,
which presumably took place at an early stage of the evolution of
the universe. One can find a discussion of the  CC problem
in the QFT framework in Ref.~\refcite{ilyasola02}.

In the MSt model, the contribution of the massive scalar fields
emerges in the effective action of gravity through the
renormalization of the cosmological constant  $\Lambda$
term in Einstein's equations. Hence, when we take $\Lambda=0$
in Eq.(\ref{eqmov}),
the MSt model is left solely with the contribution due to
the masses of the fermion fields, which is within the definition of $\tilde{f}$ [Eq.(\ref{ftil})],
the $\beta$-function which renormalizes the gravitational constant.

On the other hand, the $\beta$-function which renormalizes the $\Lambda$ term can be
linked to a dimensionless expression $\tilde{g}$  (see Eq.~(\ref{gtil}) below).
The parameter $\tilde{g}$ is given by an algebraic sum of the fourth powers of the
masses of the fermion and scalar fields, taking their statistic
and multiplicities into account. It is not obvious that this sum should cancel out at all
stages of inflation, even if the supersymmetry is initially
present. Therefore
$\tilde{g}$ may not vanish. Hence, when we take
$\Lambda\neq 0$ the contribution of the massive scalar fields
should, in principle, contribute to the solution of the MSt
model and the effect of such contributions on the
inflation should be investigated.

The stability criterion in the MSt model depends,
in principle, as we are going to show,
on the size and sign of $\tilde{g}$ and on the supersymmetry
breaking (the
value of $\Lambda$). The stability at the initial inflationary
period is one of the great successes of the MSt model.
The MSt model does not present any initial
condition problem for $\Lambda=0$ since the stability criterion is
satisfied.\cite{asta}
In this paper we reconsider the solutions and the stability condition
for the MSt model assuming the minimum supersymmetric standard
model (MSSM) particle content {at the beginning of inflation}
and the natural values of
$\tilde{g}$ and $\Lambda$ at  the corresponding energy
scale. We show here that, in contrast to the naive expectations, a non-vanishing CC and $\tilde{g}$
(the last depending of multiplicities and masses of the fermion
and scalar constituents of the supersymmetric model)
do not destroy the stability in the MSt model.

The paper is organized as follows. In section \ref{sec2}, we
present the framework of the modified Starobinsky (MSt) model,
revisiting the stability criterion for a non-vanishing $\Lambda$ and $\tilde{g}$  as well as
the approximate analytic solution for the MSt model in section \ref{sec3}.
We introduce the natural values of the parameters to analyze the
solutions and stability criterion numerically in section \ref{sec4}.
The conclusions of the paper are presented in section \ref{sec5}.

\section{The framework of the modified Starobinsky model\label{sec2}}

In this section, we discuss the anomaly induced inflation formalism
of the Starobinsky model\cite{star} following the notations in
Refs.~\refcite{anju}-\refcite{massivecase}.
In the Starobinsky model, inflation comes from the contribution
of the quantum effects of massless matter (scalar, fermion and
vector) fields.\cite{star,anju,wave} Assuming a non-minimal
conformal coupling between the scalar field and gravity,
$\xi=1/6$, the massless matter fields are invariant under the
local conformal transformation of the fields and the metric
\begin{equation}
g_{\mu\nu} \to {\bar g}_{\mu\nu}\,e^{2\sigma(\eta)}\,,
\label{transfmetric}
\end{equation}
where $\sigma(\eta)=\ln{a(\eta)}$, $dt=a d\eta$, \ $\eta$ is the
proper (conformal) time and $a$ is the scale factor.

The massless matter action satisfies the conformal Noether
identity which implies that the energy-momentum tensor is traceless at the classical
level. However, the one-loop quantum
contribution of the massless matter fields cause a trace anomaly
$<T> \neq 0$ which is useful to calculate the induced effective
action.\cite{book,rie,frts} In this conformal anomaly method, the
anomaly-induced effective action $\Ga_{ind}$ is derived from the
trace anomaly using  the conformal metric (\ref{transfmetric}).
We assume the spatially flat ($k=0$) line element of the
Friedmann-Lema\^itre-Robertson-Walker (FLRW) metric
\begin{equation} ds^2= a^2(\eta) ( d\eta^2 - d{\bf x}^2)\,.
\label{coord}
\end{equation}
General equations for
$k=\pm 1$ (non flat space) can be found in Refs.~\refcite{star,wave}
for the massless theory and for the
modified Starobinsky (MSt) model in Ref.~\refcite{asta}.

In the MSt model, inflation starts with a massive supersymmetric
(SUSY) particle content $N_{0,1/2,1}$, where $0,1/2$ and $1$
correspond to scalar, fermion and vector fields, respectively. The
theory is not conformal invariant anymore due to the masses of the scalar
and fermion fields. However, applying a conformal description,\cite{cosmon1}
the massive theory becomes conformal invariant at the
classical level and we can use the conformal anomaly method to
derive the effective action.\cite{massivecase,asta}

The classical vacuum action which
provides the possibility to renormalize the massive theory, in its
minimal version contains, besides the Newton $G$ and $\Lambda$
terms, the parameters $\,a_1,\,a_2,\, a_3$, defined according to
$$
S_{vac} \,\,=\,\,S_{HD}\,+\,S_{EH}\,,
$$
where $S_{HD}$ is the part which contains higher derivatives of
the metric\footnote{One has to notice that the introduction of the
non-conformal terms like $\,\int\sqrt{-g}R^2\,$ is possible, but not necessary, for the
renormalization of the free conformal invariant theories (see e.g. Ref.\refcite{anju} and references therein).}
\begin{equation} S_{HD} = \int d^4x\sqrt{-g}\,\left\{ a_1 C^2 + a_2 E +
a_3 {\Box} R \right\}\,, \label{vacuumaction}
\end{equation}
where
$
C^2 = C_{\mu\nu\al\be}C^{\mu\nu\al\be} =
R_{\mu\nu\al\be}R^{\mu\nu\al\be} - 2 \,R_{\al\be}R^{\al\be} +
\frac13\,R^2\,
$
and 
$
E = R_{\mu\nu\al\be}R^{\mu\nu\al\be} - 4 \,R_{\al\be}R^{\al\be} +
R^2
$
are the square of the Weyl tensor and the integrand of the
Gauss-Bonnet topological term, respectively,
and $S_{EH}$ is the Einstein-Hilbert action
\begin{equation}
S_{EH}\, =\, -\,\frac{1}{16\pi G}\,\int d^4x\sqrt{-g}\,(R +
2\La)\,. \label{Einstein}
\end{equation}

Using the method proposed in Refs.~\refcite{massivecase,asta}, the
anomaly-induced effective action $\Ga_{ind}$ can be derived from
the trace anomaly of the massive theory with the usual
conformal transformations to the fields and the
conformal metric (\ref{transfmetric}).
We obtain the total effective action adding the classical terms
$$
\Gamma \,\,\cong \,\,S_{HD} \,+\, \int d^4
x\sqrt{-{\bar g}} \,\{w{\bar C}^2\sigma \,+\, b\,({\bar E}
-\frac23 {\bar \nabla}^2 {\bar R})\,\sigma \,+\, 2 b\,\sigma{\bar
\Delta}\sigma \}
$$
$$
 \,-\, \frac{3c+2b}{36}\,\int
d^4x\sqrt{-g}\,R^2\,-\, \int d^4 x\sqrt{-{\bar
g}}\,e^{4\si}\,\cdot\, \Big[\frac{\La}{8\pi
G}\,-\,g\cdot\si\,\Big]
$$
\begin{equation} -\,\int d^4 x\sqrt{-{\bar g}} \,e^{2\si}\,[{\bar
R}+6({\bar \na}\si)^2] \,\cdot\,\Big[\, \frac{1}{16\pi G} -
f\cdot\si\,\Big] \,, \label{quantum for massive}
\end{equation}
where we have discarded a possible unknown term which comes from the
integration of $\Gamma_{ind}$.
The coefficients $w,\,b,\,c,\,f$ and $\,g$ are the
$\beta$-functions for the parameters of the classical vacuum
action $\,a_1,\,a_3,a_3,\,G\,$ and $\,\La\,$, respectively. The
explicit form of these functions is:
\begin{equation}\left(
  \begin{array}{c}
    w \\
    b \\
    c \\
  \end{array}
\right) \,=\, \frac{1}{(4\pi)^2}\left(
 \begin{array}{c}
  \frac{N_0}{120} + \frac{N_{1/2}}{\,\,20} + \frac{N_1}{10}
\\
 -\frac{N_0}{360} - \frac{11N_{1/2}\,}{\,\,360}- \frac{31\,N_1}{180}
\\
  \frac{N_0}{180} + \frac{N_{1/2}}{30} - \frac{N_1}{10}
\\
\end{array} \right)\,,
\label{betafunctions}
\end{equation}
\begin{equation}
f\,=\,\frac{1}{3(4\pi)^2}\,\sum_{f}\,{N_f\,m_f^2}\,,
\end{equation}
\begin{equation} g\,=\,\frac{1}{2(4\pi)^2}\,\sum_{s} \,{N_s\,m_s^4}
-\frac{2}{(4\pi)^2}\sum_{f}\,{N_f\,m_f^4}\,. \label{fg}
\end{equation}
Here $N_f$ and $N_S$ are the multiplicities of the fermion and
scalar massive fields with the masses $m_f$ and $m_S$.
Taking the
minimal variation of Eq.(\ref{quantum for massive}) with respect
to the conformal factor $\sigma$, we obtain the following equation of motion \[
{\stackrel{....}{\sigma }{(t)}}+7{\stackrel{...}{\sigma }{(t)}}
{\stackrel{.}{\sigma }{(t)}}+4\,{{\stackrel{..}{\sigma}^{2}(t)}}
+4\,\Big( 3-\,\frac{b}{c}\Big) \,{\stackrel{..}{\sigma
}{(t)}}{{\stackrel{.}{\sigma}^{2}{(t)}}}
-4\,\frac{b}{c}\,{{\stackrel{.}{\sigma}^{4}{(t)}}} \,\, \]
\[ \,- \,\frac{M_{\rm Pl}^2}{c}\,\left[ \,\left(
{\stackrel{..}{\sigma}{(t)}}\,+2{{\stackrel{.}{\sigma}^{2}{(t)}}}
\right)\cdot
\left(1-\tilde{f}\sigma{(t)}\right)-\frac12\,\tilde{f}
\stackrel{.}{\sigma}^2{(t)}\,\right] \]
\begin{eqnarray}
+ \,\, \frac{2\,{\Lambda}M_{\rm Pl}^2}{3\, c} \,\left(1 -
\tilde{g}\sigma{(t)}-\frac{\tilde{g}}{4}\right)\,\,=\,0\,,
\label{eqmov}
\end{eqnarray}
where $M_{\rm Pl} = 1/\sqrt{8\pi G} = 2.44 \times 10^{18}\, {\rm
GeV}$ is the reduced Planck mass.
The parameters $\tilde{f}$ and $\tilde{g}$ are defined as
dimensionless functions of the previous $f$ and $g$
\begin{equation}
\tilde{f} = (16\pi G)\, f = \frac{2}{3{(4 \pi)}^2}
\,\sum_{f}\,\frac{N_f\,m_f^2}{M^2_{\rm Pl}}\,,
\label{ftil}
\end{equation}
\begin{equation}
\tilde{g} = \frac{g }{\Lambda/(8\pi G)}\,
\,=\,\frac{1}{2(4\pi)^2}\,\sum_{s}
\,\frac{N_s\,m_s^4}{{\Lambda}_{\rm ad}\,M^4_{\rm Pl}}
-\frac{2}{(4\pi)^2} \,\sum_{f}\,\frac{N_f\,m_f^4}{{\Lambda}_{\rm
ad}\,M^4_{\rm Pl}}\,, \label{gtil}
\end{equation}
where we have introduced here a dimensionless vacuum parameter, according to
$\rho_{\rm vac}= \Lambda/(8\pi G)$, as
\begin{equation}
\Lambda_{\rm ad}\equiv \rho_{\rm vac}/M_{\rm Pl}^4
\label{lambaddef}\,.
\end{equation}
The Starobinsky inflationary solution can be found assuming the
massless matter fields ($\tilde{f}=\tilde{g}=0$) and $\Lambda=0$
in Eq.(\ref{eqmov})
\begin{equation}
a(t)\equiv e^{\sigma_S(t)}=e^{H_{S} t}\,,
\label{deSitter}
\end{equation}
where $H_S \,\equiv\,H_{a}\,M_{\rm Pl}$
is the Hubble parameter in the original Starobinsky model\cite{star}
and $H_{a}=1/\sqrt{-2\,b}$ is a dimensionless constant.
The parameter $H_a$ depends on the particle content
$N_0$, $N_{1/2}$ and $N_1$ according to Eq.(\ref{betafunctions}).
A solution for $\Lambda\neq 0$ can be found in Ref.~\refcite{wave}.

When massive field contributions are taken
into account, the equation of motion (\ref{eqmov}) becomes more
complicated and can not be solved analytically. Numerical
calculations with $\Lambda=0$ have shown that $H$ is slowly
decreasing in time.\cite{massivecase,asta}
It is important that the corresponding
solution is stable until the transition point, where
heavy s-particles decouple,
supersymmetry breaks down and
another phase of inflation starts.\cite{graceexit} During
the stable
period one can find an accurate approximate solution which
closely reproduces the numerical solution. In the
next two sections we shall generalize these results by taking
the cosmological constant and its running into account.

\section{The stability conditions and the approximate solution
\label{sec3}}

In the massless theory,
the stability condition is well-known: $c > 0$.\cite{star,asta}
This condition corresponds to the assymptotic stability of
the de Sitter solution under small perturbations of the
conformal factor of the metric
$\sigma(\tau)\rightarrow \sigma_S(\tau) + y(\tau)$,
where $\sigma_S$ is given by Eq.(\ref{deSitter}).
According to
Eq.(\ref{betafunctions}), this condition corresponds to
the following inequality for the particle content of the theory
\begin{equation}
N_1 < \frac{1}{3}N_{1/2} + \frac{1}{18}N_0\,, \label{stabstar}
\end{equation}
where $N_1,\,N_{1/2},\, N_0$ are the numbers of particles with
the corresponding spin.

Furthermore, the stability condition for the massive theory,
which
was obtained in Ref.~\refcite{asta}, has the following form
\begin{equation}
3{\tilde f}{\tilde H}^2(\tau)-{\,{\tilde g}\,{\tilde
\La}(\tau)}\,>\,0\,, \label{key}
\end{equation}
with
\begin{equation}
 {\tilde H}^2(\tau)=-\frac{{\tilde M}_{\rm Pl}^2(\tau)}{4 b}\,
\Big[\,1\,+\, \Big( 1+\frac{8 b}{3}\, \frac{{\tilde
\Lambda}(\tau)}{{\tilde M}_{\rm
Pl}^2(\tau)}\Big)^{1/2}\Big]\,,\label{Htilde}
\end{equation}
where we have normalized the time to $\tau\equiv {t}/{t_{\rm Pl}}$
($t_{\rm Pl}\simeq 5.3\times 10^{-44} sec$ is the Planck time) and
\begin{equation}
{\tilde M}_{\rm Pl}^2(\tau) = M_{\rm Pl}^2\,\Big[1 -
\tilde{f}\,\sigma(\tau)\Big]
\,,\,\,\,\,\,\,\,\,\,\,\,\,\,\,\,\,\,\,\,\,\,{\tilde \La}(\tau) = \La
\Big\{1 - \tilde{g} \,\sigma(\tau)-\tilde{g}/4\Big\}\,.
\label{replace1}
\end{equation}
During inflation, the solution of the MSt model is stable (obeying
the criterion of stability) until a characteristic scale
$M_*$.\cite{asta}
After this scale, or the dimensionless scale
$\mu\equiv M_*/M_{\rm Pl}$, the Hubble
parameter decreases, becoming constant and very small
$\tilde H_*=
\dot{\sigma}(\tau)\approx \mu$, when the sparticles
decouple and the matter content $N_{0,1/2,1}$ becomes modified.
As a result, the inequality in Eq.(\ref{stabstar}) changes sign and
the universe enters into an
unstable inflation regime with an eventual transition to the FRW evolution.
We have shown that the last transition of the inflation, satisfying
the stability condition (\ref{stabstar}), is
independent of the value of the cosmological constant $\Lambda$
or the curvature $k$ in Ref.~\refcite{asta}.

Our purpose here is to consider the MSt model assuming the
possible contribution of the cosmological constant $\Lambda$
and $\tilde{g}$ to the numerical solution.
The approximate solution for the MSt model is obtained assuming
$\tilde{f}$ small,
a slowly varying Hubble parameter and, in addition,
$\Lambda=0$.\cite{massivecase,asta}
Discarding the term $\tilde{f}\stackrel{.}{\sigma}^2$ in
Eq.(\ref{eqmov}) we obtain exactly the equation of motion
for the massless matter fields.\cite{anju,wave}
The approximate analytic solution is then obtained taking
$\Lambda=0$ in Eq.(\ref{replace1}). Thus, according
to the first expression of Eq.(\ref{replace1}), integrating
the running Hubble parameter
$H(\tau)=\,{\sigma}^{\prime}(\tau)\equiv H_a\,
{\tilde M}_{\rm Pl}(\tau) $
we find\footnote{Similar behaviour arises in simple inflationary
models with potential
$V\propto\phi^2$ (see Ref. \refcite{star78}) and in the limiting
form of the original Starobinsky model, called ``quasi-de Sitter"
stage in Ref. \refcite{star83}. Therefore, from the theoretical
point of view, the advantage of the MSt model is that it is
completely free of any ad hoc assumption (such as inflaton
potential) and it starts with a very stable inflationary
solution without any fine tuning of the parameters. In
the massless theory, for instance, it is necessary to
introduce some additional term or terms into the
classical action of vacuum (e.g. a sufficiently large
positive coefficient in the $(\int \sqrt{-g}R^2)$-term)
to provide $c>0$.}
\begin{equation}
\sigma(\tau)\,=\,H_{a}\,\tau
\,-\frac{H_{a}^2\tilde{f}}{4}\,\tau^2\,.\label{parabola}
\end{equation}
This approximate analytic solution can be
understood as the massive contribution (with $\tilde{f}\neq 0$
and $\Lambda=0$) arises as a running G, $G(\tau)\equiv [8\pi {\tilde
M}_{\rm Pl}^2(\tau)]^{-1}$, to the Starobinsky model.
The approximate solution Eq.(\ref{parabola})
shows a very good agreement with the numerical solution of the
MSt model for the case $\Lambda=0$.\cite{asta}
However, let us remark that the corresponding analysis
was performed for the particular case $\Lambda=0$ and $\tilde{g}=0$.
In particular, we assumed $\Lambda=0$
and that our numerical solutions is stable as far as $\tilde f>0$
[Eq.~(\ref{ftil})], such that the criterion
(\ref{key}) was satisfied. However, the stability criterion
for the massive theory (\ref{key})  may depend,
in principle, on the value of $\Lambda$ and also on the
size and sign of $\tilde{g}$. The last parameter can be
different from zero, since it is not necessary that there is
an algebraic cancelation between the sum of the fourth powers
of the masses of the fermion and scalar fields.
Using a natural value for $\Lambda$ during the inflationary
period and the possible
values of $\tilde{g}$, in the next section we analyze the
solutions and the stability criterion for the
general MSt model.

\section{Numerical analysis \label{sec4}}

In this section, we discuss the natural values of the parameters $\tilde{f}$ and
$\tilde{g}$ for $\Lambda$ at the SUSY scale.
Using these values we compare the approximate parabolic solution with the full
numerical solution of the MSt model and the
stability criterion for the initial evolution of
the inflationary period.
Recent astronomical observations indicate a small
value for the cosmological constant
\begin{equation}
\Lambda_{\rm obs} \approx 0.7 \rho_c^0 \sim 10^{-47} \, {\rm
GeV}^4\sim 10^{-121}\,M_{\rm Pl}^4\,.
\end{equation}
It can be much more larger at higher energy scales, for example, on
SUSY (or GUT) scales. This difference can be due to phase transitions,
e.g. the electroweak, which took place between the
present time and the SUSY scale. Let us suppose $M_{\rm
SUSY}\sim 10^{16}\,{\rm GeV}\simeq 10^{-3} M_{\rm Pl} $ which is
consistent with the GUT scale ($\sim 10^{14}$ - $10^{16}) {\rm
GeV}$. In this case,
\begin{equation}
\rho_{\rm vac}\simeq M_{\rm SUSY}^4\sim 10^{-12} M_{\rm Pl}^4\,.
\label{susyvac}
\end{equation}
The difference between the two above values is a manifestation of
the well-known cosmological constant (CC) problem. Certainly it is much easier to assume
that the vacuum energy is zero than to solve the CC problem.\cite{rvdec3}\cdash\cite{rvdec1}
However, let us remark that almost all deviations from a
``perfect equilibrium" vacuum lead to a non-zero vacuum energy
density which is proportional to the perturbations of the vacuum.\footnote{For a recent discussion see e.g. Refs.~\refcite{volovik} and \refcite{nova}.}
Nevertheless, from Eq.(\ref{susyvac}), the natural
value of the dimensionless vacuum parameter $\Lambda_{\rm ad}$ [Eq.(\ref{lambaddef})] at the SUSY scale is
\begin{equation}
\Lambda_{\rm ad} \approx 10^{-12}\,. \label{lambvalue}
\end{equation}
For the numerical analysis, we use the
particle content $N_{1,1/2,0}=(12,48,104)$ of the minimum
supersymmetric standard
model (MSSM) and $\tilde{f}$ in the
range
\begin{equation}
\tilde{f}\sim (10^{-5}-10^{-3})\,,
\label{ftilrange}
\end{equation}
which corresponds to the scale of Eq.(\ref{susyvac}) and a
typical matter content $N_f+N_s\sim 10-1000$.
\begin{figure}
  \centering
{\includegraphics[width=90mm]{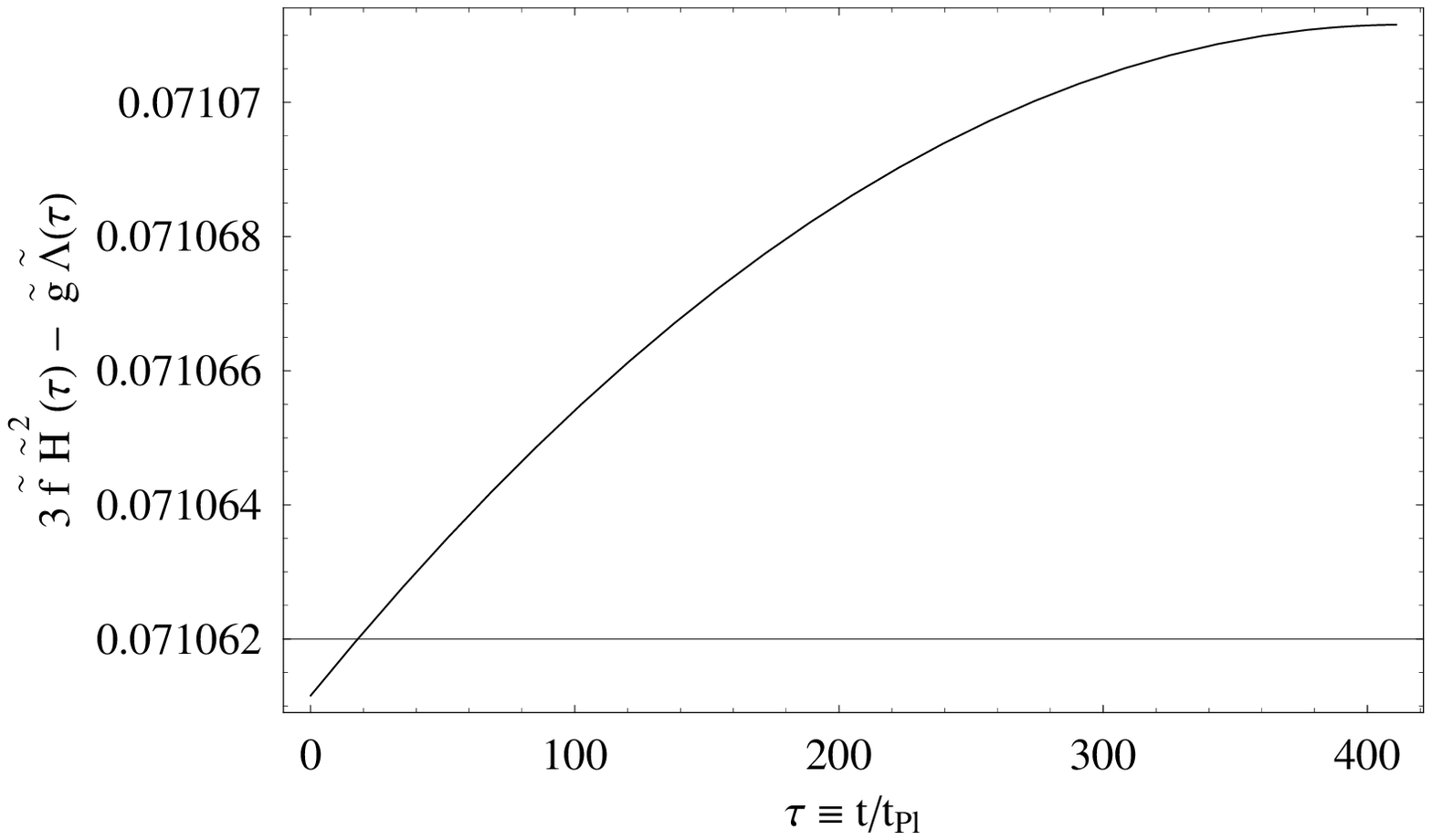}}
{\includegraphics[width=90mm]{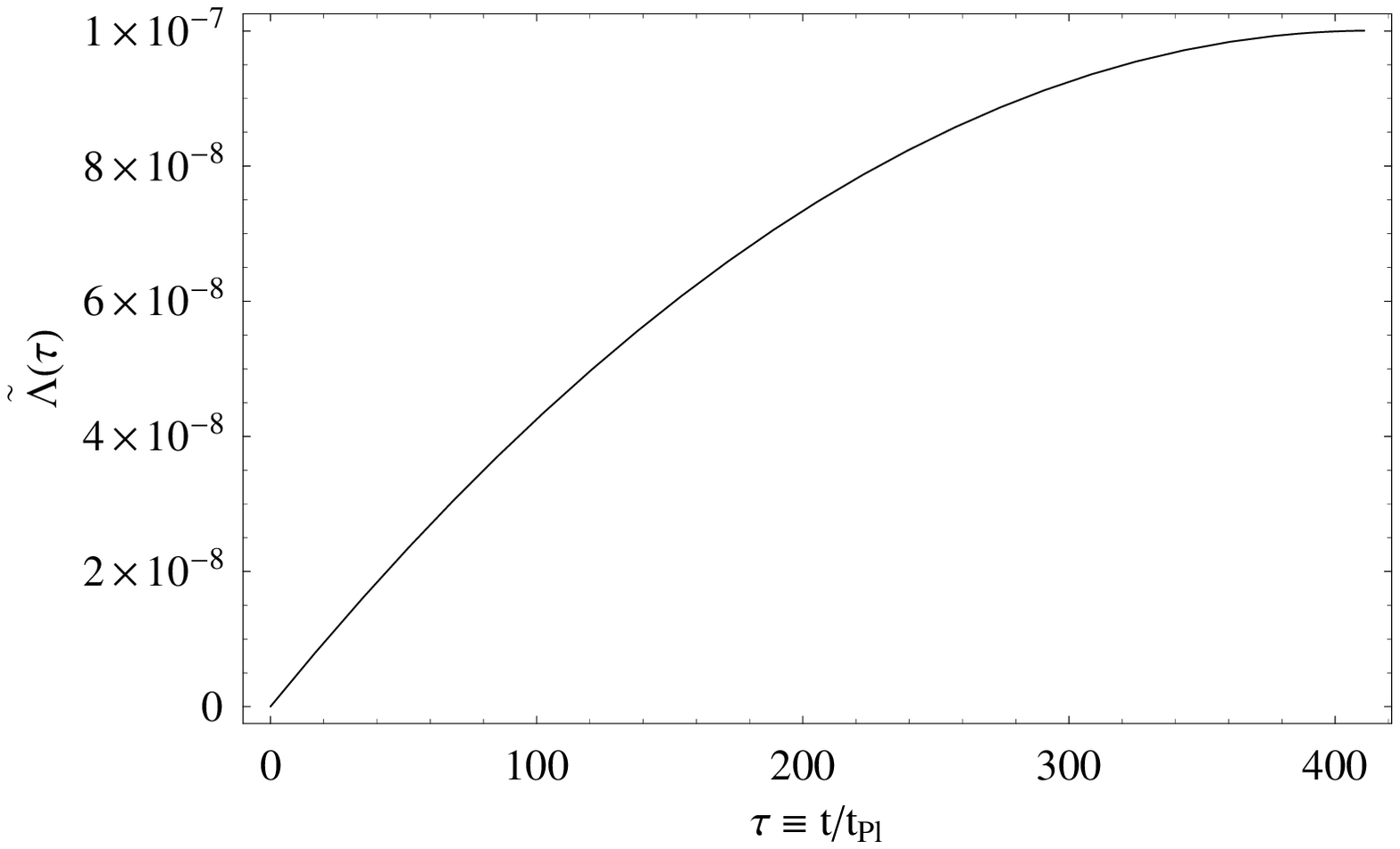}}
\caption{a) The stability criterion for the massive theory
Eq.(\ref{key}) assuming the
MSSM particle content $N_{1,1/2,0}=(12,48,104)$, the natural
value of the dimensionless $\Lambda$ at the SUSY scale
(\ref{lambvalue}) and a combination
of extreme values of the parameters
$\tilde{f}= 10^{-3}$ and $\tilde{g}=-10^{2}$. b) The time
dependence of $\tilde \Lambda(\tau)$ from the transformations
Eq.(\ref{replace1}) for the same
values of the parameters as in Fig.1a. }
\end{figure}

Whether we assume the masses of the fermions and scalars to be of the same order or not, we can
stipulate the parameter $\tilde{g}$ from Eq.(\ref{gtil}) as
\begin{equation}
\tilde{g}\sim \pm\frac{\tilde{f}}{\Lambda_{\rm ad}} \,\sum_{s}
\,\frac{\,m_s^2}{M^2_{\rm Pl}}\,.
\label{gtilap}
\end{equation}
Let us extrapolate the possible difference between
the sum of the scalar masses squared and
$\Lambda_{\rm ad}$, as well as the simplified approximation used in Eq.(\ref{gtilap}),
in the range $(10^{-3}-10^{5})$. Thus, using Eqs.(\ref{lambvalue}) and (\ref{ftilrange}) we find
\begin{equation}
\tilde{g}\sim \pm \,(10^{-8}-10^{2}) \,.
\label{gtilad}
\end{equation}

We begin the analysis solving the equation of motion of Eq.(\ref{eqmov}) for the dimensionless $\Lambda$ at the SUSY scale [Eq.(\ref{lambvalue})]
varying $\tilde{f}$ and $\tilde{g}$ within the range of Eqs.(\ref{ftilrange}) and (\ref{gtilad}), respectively. We check the stability criterion [Eq.(\ref{key})] during the inflationary period assuming the possible combinations of the range in  $\tilde{f}$ and $\tilde{g}$, as well as the possibility of positive and negative signs in $\tilde{g}$
until the end of inflation which should happens at $t_{\rm end}\sim 2/H_a \tilde{f}$, according to the approximate parabolic solution (\ref{parabola}).
We checked that the stability criterion [Eq.(\ref{key})] is satisfied for all possible combinations  of the values of the parameters.
We found that $3{\tilde f}{\tilde H}^2(\tau)-{\tilde g}\,{\tilde
\La}(\tau)$ is positive until the end of inflation and
the numerical solution is independent of the initial conditions.
As an example, we show the stability criterion for the massive theory assuming a MSSM particle content, the natural value of the dimensionless $\Lambda$ at the SUSY scale [Eq.(\ref{lambvalue})] and a combination of extreme values of the parameters
$\tilde{f}= 10^{-3}$ and $\tilde{g}=-10^{2}$ in Fig.1a. The result can be understood looking at the time dependence of $\tilde \Lambda(\tau)$ from the transformations of Eq.(\ref{replace1}). We notice that one can neglect the constant terms so that the function $\tilde \Lambda(\tau)$ has the opposite sign of $\tilde g$ (since $\sigma'(\tau)=H(\tau)>0$), or ${\tilde \La}(\tau) \propto
- \tilde{g} \,\sigma(\tau)$ as shown in Fig.1b.
This mean that the stability criterion is then $\sim
3{\tilde f}{\tilde H}^2(\tau)+{\tilde g}^2\,{\tilde
\sigma(\tau)}$ which always  has a positive sign.

\begin{figure}
  \centering
{\includegraphics[width=120mm]{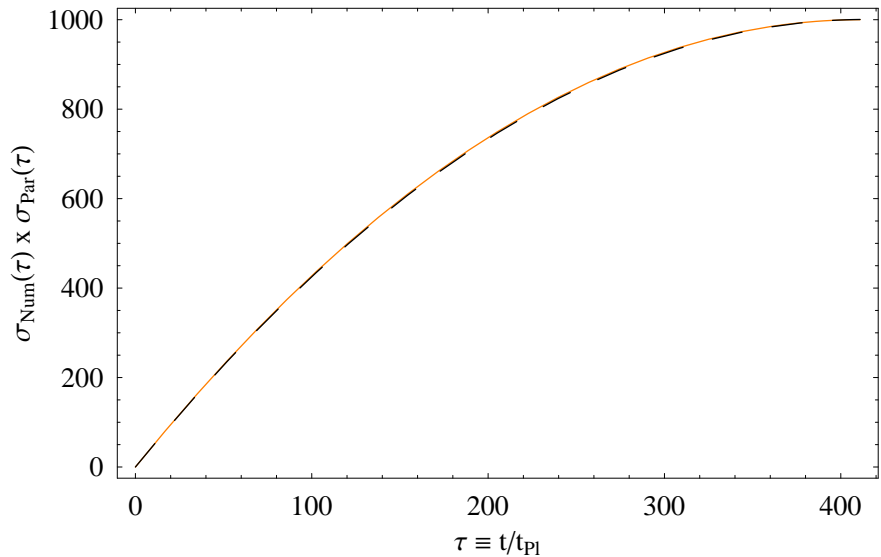}}
  \caption{The numerical solution of Eq.(\ref{eqmov}) for the same values of the parameters as in Fig.1  versus the approximate parabolic solution (the dashed line).}
\end{figure}

The numerical solution of Eq.(\ref{eqmov}) can be
compared to the approximate parabolic solution
[Eq.(\ref{parabola})] for all possible combinations of the
parameters values discussed above. As a result we found an
accurate compatibility between the numerical and approximate
solutions. We show in Fig.2 the approximate versus the
numerical solution for the MSSM particle
content with the same extreme values of the parameters used
in Fig.1.
The dashed line shows the approximate parabolic solution
[Eq.(\ref{parabola})].
The contribution of the difference due to massive fermions and bosons, which is within the definition of $\tilde{g}$
can not be distinguished in the numerical solutions. The massive fermions
are mainly responsible for slowing down $\sigma(\tau)$ and we can
safely consider $\tilde{g}$ and $\Lambda$ vanishing as an
approximation. The approximate solution is also robust in this
stage.

\section{Conclusions\label{sec5}}

We considered the inflationary solution in the modified
Starobinsky (MSt) model, taking into account the cosmological
constant $\Lambda$ and quantum contributions to the vacuum
energy density
from massive fermions and bosons. The fields with different
statistics contribute with opposite signs and the overall quantum
effect is due to the difference of their contributions.
The natural value for $\Lambda$ corresponds to the
supersymmetry breaking value at the MSSM scale. We also assumed the
corresponding particle spectrum for evaluating the quantum
contributions. The corresponding dimensionless parameter is called
$\tilde{g}$.
It turns out that the agreement between approximate analytical and
numerical solutions is robust for the natural values of the
above-mentioned parameters. Furthermore, the numerical solution is
not sensitive to the sign of $\tilde{g}$. The latter means that the
contribution of the massive fermions can be smaller or larger than
the contribution of the massive boson particles without visible
effect to the numerical solution of the model. The inflation
remains stable until the point when the s-particles decouple and
supersymmetry breaks down, for any natural choice of the
parameters.

On the other hand, the renormalization of the inverse Newton
constant, parametrized by the dimensionless quantity  $\tilde{f}$,
is very important. This parameter is always positive and its
magnitude depends only on the fermion spectrum of the theory.
Indeed, $\tilde{f}$ is responsible for decreasing the value of the
Hubble parameter in the course of inflation.

Finally, it is very important that the value of the cosmological
constant $\Lambda$ and the contributions of the massive
particles do not destroy the stability which holds at the initial
stage of inflation. Hence, the MSt model provides a possibility to
describe all stages of inflation without fine-tuning of the
parameters of the theory and/or initial data.

\vskip 8mm

\noindent {\bf Acknowledgments.} I am very grateful to I. L.
Shapiro and R. Opher for numerous discussions and helpful comments on the
manuscript and to J. Sol\`a for useful discussions. This work has
been partially supported by MEC and FEDER under
project FPA2007-66665 and by DURSI Generalitat de Catalunya
under project 2005SGR00564 and in part also by the Brazilian agency FAPESP (grants 2003/04516-0 and 2006/56213-9).

\end{document}